\documentclass{jaa}
\usepackage{hyperref}

\usepackage{natbib}
\usepackage{graphicx}

\begin{document}\sloppy

\title{H{\sc I} line analysis of Herbig Ae/Be stars using X-Shooter spectra}


\author{B. Shridharan \textsuperscript{1,*}, Blesson Mathew\textsuperscript{1}, R. Arun\textsuperscript{2}, Cysil Tom Baby\textsuperscript{1}}

\affilOne{\textsuperscript{1}CHRIST (Deemed to be University), Bangalore  \\}

\affilTwo{\textsuperscript{2}Indian Institute of Astrophysics, Bangalore}


\twocolumn[{

\maketitle

\corres{shridharan.b@res.christuniversity.in}

\msinfo{04 November 2022}{18 March 2023}

\begin{abstract}
Herbig Ae/Be stars are intermediate-mass pre-main sequence stars undergoing accretion through their circumstellar disk. The optical and infrared (IR) spectra of HAeBe stars show H{\sc I} emission lines belonging to Balmer, Paschen and Brackett series. We use the archival X-Shooter spectra available for 109 HAeBe stars from Vioque et al. (2018) and analyse the various H{\sc I} lines present in them. We segregated the stars into different classes based on the presence of higher-order lines in different H{\sc I} series. We discuss the dependence of the appearance of higher-order lines on the stellar parameters. We find that most massive and younger stars show all the higher-order lines in emission. The stars showing only lower-order lines have $T_{eff} <$ 12000 K and an age range of 5-10 Myr. We perform a Case B line ratio analysis for a sub-sample of stars showing most of the H{\sc I} lines in emission. We note that all but four stars belonging to the sub-sample show lower H{\sc I} line ratios than theoretical values, owing to the emitting medium being optically thick. The H{\sc I}~line flux ratios do not depend on the star's spectral type. Further, from the line ratios of lower-order lines and Paschen higher-order lines, we note that line ratios of most HAeBe stars match with electron density value in the range $10^9 - 10^{11} cm^{-3}$. The electron temperature, however, could not be ascertained with confidence using the line ratios studied in this work. 
\end{abstract}

\keywords{spectroscopy---emission-line stars---Case B recombination.}

}]


\doinum{12.3456/s78910-011-012-3}
\artcitid{\#\#\#\#}
\volnum{000}
\year{0000}
\pgrange{1--}
\setcounter{page}{1}
\lp{1}

\section{Introduction}

Herbig Ae/Be (HAeBe) stars are pre-main sequence (PMS) stars with masses ranging from 2 to 10 solar masses (M$_\odot$). They show several emission lines in their optical and infrared (IR) spectrum, which are known to arise through various formation mechanisms, i.e., accretion column, circumstellar disk and/or winds \citep{Muzerolle_2004, cabrit1990forbidden, kurosawa2006}. They also exhibit IR excess, indicating the thermal heating and reprocessing from the dust present in the circumstellar environment of the star \citep{hillenbrand1992herbig, malfait1998ultraviolet}. The optical/IR emission lines belonging to the H{\sc I} series, namely, Balmer, Paschen and Brackett lines, are often seen in Young Stellar Objects (YSOs) spectra and are known to arise from the material accretion from the disk. They also arise from stellar winds and jets of the most active YSOs.

The current understanding of the accretion process in HAeBe is the magnetospheric accretion model. Although initially introduced to explain accretion in highly magnetic low-mass PMS known as T-Tauri stars (TTS, \citealp{hartmann1994magnetospheric, muzerolle2001emission}), studies have shown that it can be extended till late HBe stars. Since massive HBe stars do not possess convective outer layers, the magnetic fields are weak to sustain accretion. Hence for HBe stars, the accretion may predominantly happen through boundary-layer accretion \citep{mendigutia2020mass}. A strong correlation exists between accretion luminosity and H{\sc I} emission line luminosity in YSOs \citep{muzerolle1998brgamma, calvet2004mass, natta2006accretion, alcala2014x}. The mass accretion rates are derived from H{\sc I} emission lines, especially H$\alpha$ and Br$\gamma$ \citep{arun2019mass, wichittanakom2020accretion, grant2022tracing}, assuming the magnetospheric infall model. Given the complex circumstellar environment of HAeBe stars, the H{\sc I} emission can arise from diverse physical conditions. This contribution from various regions makes it difficult to study the optical H{\sc I} lines, often affected by opacity effects. The higher order H{\sc I} lines belonging to Brackett and, to some extent, the Paschen series are less complicated and have lower optical depths \citep{folha2001near}. Hence, simultaneous analysis of emission lines belonging to different H{\sc I} series is required.

\begin{figure*}
    \centering
    \includegraphics[width=2\columnwidth]{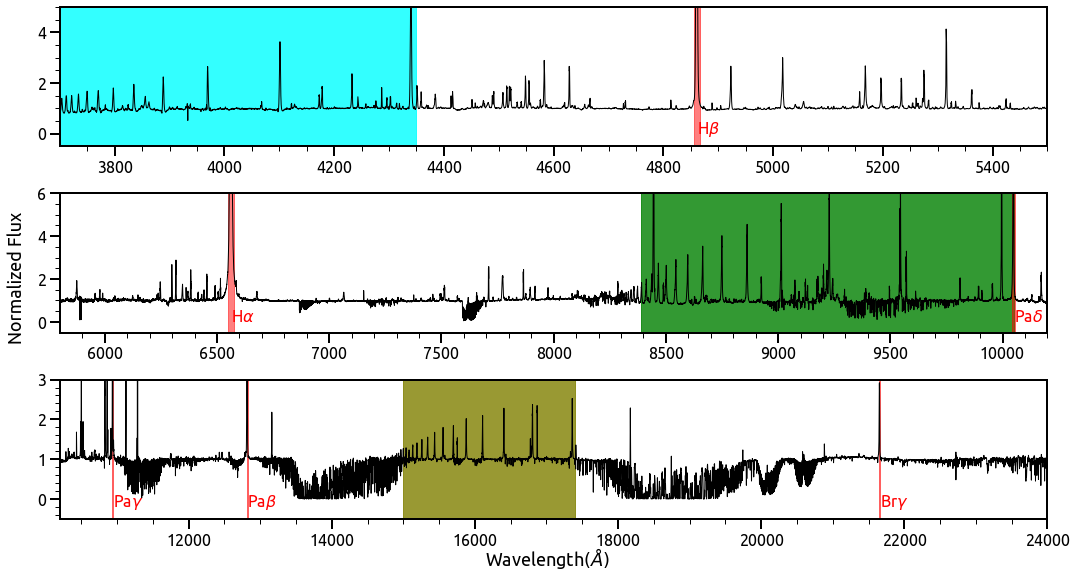}
    \caption{A representative continuum normalized spectrum in which all the higher order lines in emission are shown. The three panels shown denote the three different arms of the X-Shooter (UVB, VIS and NIR). The important lower-order lines are marked and annotated in red. The higher-order line regions of Balmer, Paschen and Brackett are highlighted in cyan, green and olive shades respectively.}
    \label{fig:spectra}
\end{figure*}

The Case B recombination model, in which the gas is optically thick to Lyman series and continuum photons, can be considered a close approximation to the H{\sc I} line emitting regions in YSOs. It has been previously used to explain the H{\sc I} decrement values \citep{nisini2004observations, bary2008quiescent, kraus2012nature, whelan2014accretion}. Many of these works only study the line ratios of lower-order emission lines (such as H$\alpha$, Br$\gamma$). It is crucial to study the higher-order lines of the Brackett and Paschen series to disentangle the opacity effects. \citet{benedettini1998iso} found that the observed ratios do not match expected Case B models or a completely ionized wind model by studying the IR H{\sc I} recombination lines in two HAeBe stars. \citet{nisini2004observations} studied the Brackett decrement of a class I source and found that the Br lines are optically thick upto higher order lines when compared to Case B ratios for electron Temperature (T) = 10000 K, electron density (log(n$_e$\footnote{n$_e$ given in $cm{^{-3}}$ units})) = 6 and T = 6000 K, log(n$_e$) = 4. For TTS, \citet{bary2008quiescent} studied the IR H{\sc I} lines using low-resolution spectra of 16 TTSs. They arrived at the result that the Paschen and Balmer decrement values fitted best with physical conditions of the emitting medium having T $<$ 2000 K and log(n$_e$) = 10, which is in contrast to the temperatures expected in magnetospheric columns (6000--12000K; \citealp{muzerolle2001emission}). However, a similar analysis for a sample of HAeBe stars in the literature is lacking. 

Interferometric analysis of HAeBe stars by \citet{eisner2009spatially} found that the formation of Br$\gamma$ emission is from a compact region around the star, consistent with the magnetospheric accretion paradigm. The above result was in contrast to the findings of \citet{kraus2008detection}, where observation of only one HAeBe was consistent with funnel flow. Further, \citet{beck2010spatially} found many TTSs with spatially extended Br$\gamma$ emission from the point source. It is clear that the H{\sc I} line emission (especially Br$\gamma$) can arise from different regions. Hence, we do not attempt to disentangle the physical conditions of different line formation regions. 

In this work, we use the available X-Shooter spectra of HAeBe stars, which provide near-simultaneous optical/NIR spectra in the wavelength range (3500--25000 $\AA$). We then classify the stars based on the presence of emission lines belonging to the Balmer, Paschen and Brackett series. We check for a correlation between the presence of higher-order lines and the stellar parameters of HAeBe stars. Then, we calculate the line flux ratios from the measured line equivalent widths (EWs) and continuum flux values from Gaia DR3 synthetic photometry \citep{gaiamainpaper2022, gaiasynth} for stars showing all higher-order lines in emission. We compare various line decrements to the Case B recombination values, provided in \citet{storey1995recombination}, to understand the physical conditions of the gaseous medium.     

Section 2 briefly describes the X-Shooter instrument and the HAeBe sample used for the present study. Section 3.1 provides an account of the relation between the presence of higher-order emission lines and the stellar parameters. Section 3.2 describes the Case B recombination analysis performed for a subset of HAeBe stars showing higher-order hydrogen emission lines in their spectra. Section 4 provides a summary of the results of this work.

\section{Data}

The most extensive compilation of HAeBe stars, which has 252 stars, is by \citet{vioque2018gaia}. The catalogue provides accurate stellar and disk parameters of the HAeBe stars estimated using Gaia DR2 data. \citet{fairlamb2016vizier} has done an extended analysis of accretion-related properties of HAeBe stars for 91 HAeBe stars using X-Shooter spectra. X-Shooter \citep{xshooter2011} provides near-simultaneous wavelength coverage from 3000 to 25000 $\AA$ over three separate arms: UVB:3000–5600 $\AA$; VIS:5500–10,200 $\AA$ and NIR:10200–24800 $\AA$. The resolution of the instrument varies between 8000-12000 depending on the wavelength, and the slit size used. We crossmatched the coordinates of 252 HAeBe stars with those listed in ESO Phase 3 archive\footnote{\url{http://archive.eso.org/wdb/wdb/adp/phase3_spectral/form}}, with a search radius of 20". However, only spectra within 3" of the central co-ordinate of the star were retained to avoid source confusion. We found that 109 HAeBe stars have spectra in the X-Shooter archive.  The EW values used in this work are from \citet{fairlamb2016vizier}, where the observed EW values are corrected for underlying photospheric absorption and veiling carefully after taking care of the telluric bands. The photospheric templates corresponding to the stellar parameters of the star were taken from \citet{munari2005}. The downloaded spectra were handled and visually checked using the Image Reduction and Analysis Facility (IRAF; \citealp{tody1986}).

\begin{figure*}
    \centering
    \includegraphics[width=2\columnwidth]{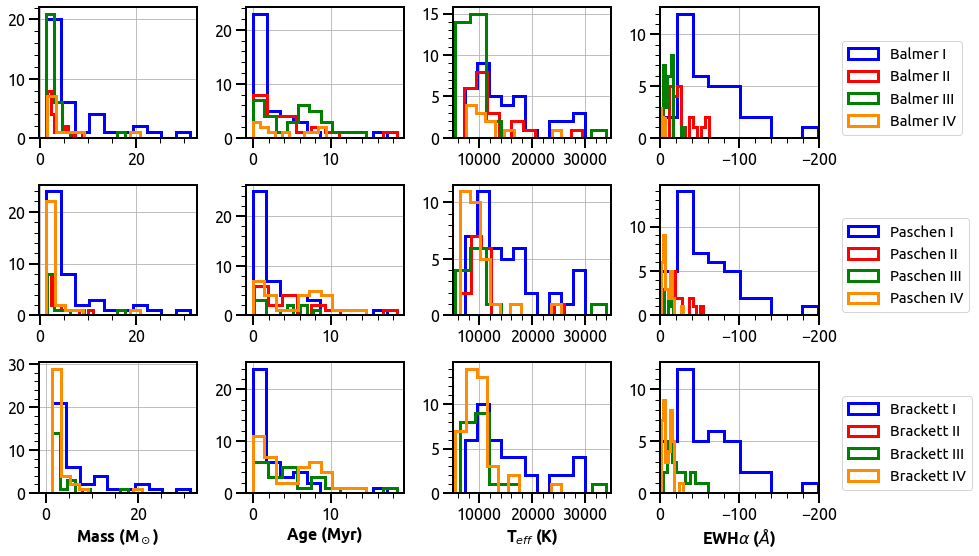}
    \caption{Histogram representations of different groups of H{\sc I} series as explained in Section 3.1. The panels show the relationship between different groups and the stellar parameters estimated by \citet{vioque2018gaia}. Series I, II, III and IV are given in blue, red, green and orange, respectively. }
    \label{fig:hists}
\end{figure*}

\section{Analysis and Results}

The presence of H{\sc I} lines belonging to the Balmer, Paschen and Brackett series in each star was visually identified. The spectra are then segregated into groups based on the presence of higher-order lines in each series. The distribution of stellar parameters for the sample of stars is taken from \citet{vioque2018gaia}. The EW values taken from \citet{fairlamb2016vizier} were converted into line flux values based on the continuum flux values using the synthetic photometry magnitudes provided in the latest Gaia DR3 release \citep{gaiasynth}. We used `pyneb'\footnote{\url{http://research.iac.es/proyecto/PyNeb//}} python library to retrieve the theoretical Case B recombination values calculated by \citet{storey1995recombination}.

\subsection{Classification of HAeBe stars based on the presence of H{\sc I} lines}

The presence of different series of H{\sc I} emission lines depends on the physical conditions of the emitting region. It is known that HAeBe stars have a complex gas and dusty disk structure. Hence it is essential to study the statistical presence of various H{\sc I} emission lines with respect to the stellar parameters.

The spectra were grouped based on the presence of series lines, as explained below. 
\begin{itemize}
    \item Series I -- If most of the higher-order lines are seen in emission
    \item Series II -- If only the lower order of the series is seen in emission (H$\alpha$ and H$\beta$ in the case of Balmer; Pa$\beta$ and Pa$\gamma$ in the case of Paschen)
    \item Series III -- If only the lowest order line were seen in emission (only H$\alpha$ in case of Balmer; only Pa$\beta$ in case of Paschen; only Br$\gamma$ in case of Brackett)
    \item Series IV -- No emission lines in the series were observed
\end{itemize}

\begin{table}[]
\resizebox{\columnwidth}{!}{%
\begin{tabular}{|l|l|l|l|}
\hline
\footnotesize{Series} & \footnotesize{Balmer} & \footnotesize{Paschen} & \footnotesize{Brackett}\\ \hline
\footnotesize{I} & \footnotesize{36}\%   & \footnotesize{46}\%    & \footnotesize{40}\%     \\ \hline
\footnotesize{II}    & \footnotesize{21}\%   & \footnotesize{12}\%    & \footnotesize{-}    \\ \hline
\footnotesize{II}I   & \footnotesize{31}\%   & \footnotesize{13}\%    & \footnotesize{40}\%     \\ \hline
\footnotesize{IV}    & \footnotesize{11}\%   & \footnotesize{29}\%    & \footnotesize{20}\%     \\ \hline
\end{tabular}%
}
\caption{Percentage distribution of different groups classified based on the presence of different series of H{\sc I} emission lines. }
\label{tab:my-table}
\end{table}

 Table \ref{tab:my-table} shows the statistical breakup of various groups for HAeBe in our sample. It also shows that the higher-order lines of the Paschen and Brackett series are present in a higher percentage of stars than the Balmer series. Also, Paschen II and III stars are only ~12\% and ~13\%, while 40\% of stars belong to Brackett III. 

Figure \ref{fig:hists} shows a histogram representation of various stellar parameters taken from \citet{vioque2018gaia} with respect to the different classes defined. The inferences derived from this distribution analysis are listed below,
\begin{itemize}
    \item We found that most HBe stars belong to `Series I' due to the presence of higher-order lines in emission. This correlation may suggest a distinctness in the circumstellar environment of HBe stars compared to HAe stars.
    \item The age dependence of `Series I' stars is evident from the histograms, where most of the `Series I' stars are young, with ages less than 2 Myr. The `Series II and III' stars seem to be distributed in the 5-10 Myr age range. It is also surprising to note that some `Series II and III' have an age range of 1-5 Myr, which points to the fact that the emitting region can evolve in different ways in the initial few Myrs. 
    \item Even though spectral type and mass are related, some exciting inferences can be taken from $T_{eff}$ distribution. The `Series I' stars seem evenly distributed in $T_{eff}$ range 10000-30000K. Most stars with $T_{eff}$ $<$ 12000K seem to belong to `Series II and III'.
    \item  From EWH$\alpha$ histograms, we see that `Series I' stars are intense H$\alpha$ emitters ($> 20 \AA$). The `Balmer' groups follow the expected trend that `Balmer II' stars are slightly more intense emitters than `Balmer III' stars. Even though we did not find H$\alpha$ in emission for `Balmer IV' stars, the EWH$\alpha$ reported by \citet{vioque2018gaia} is between 0-10 $\AA$, which may be because of variability.
\end{itemize}

 \begin{figure*}
    \centering
    \includegraphics[height=8cm, width=2\columnwidth]{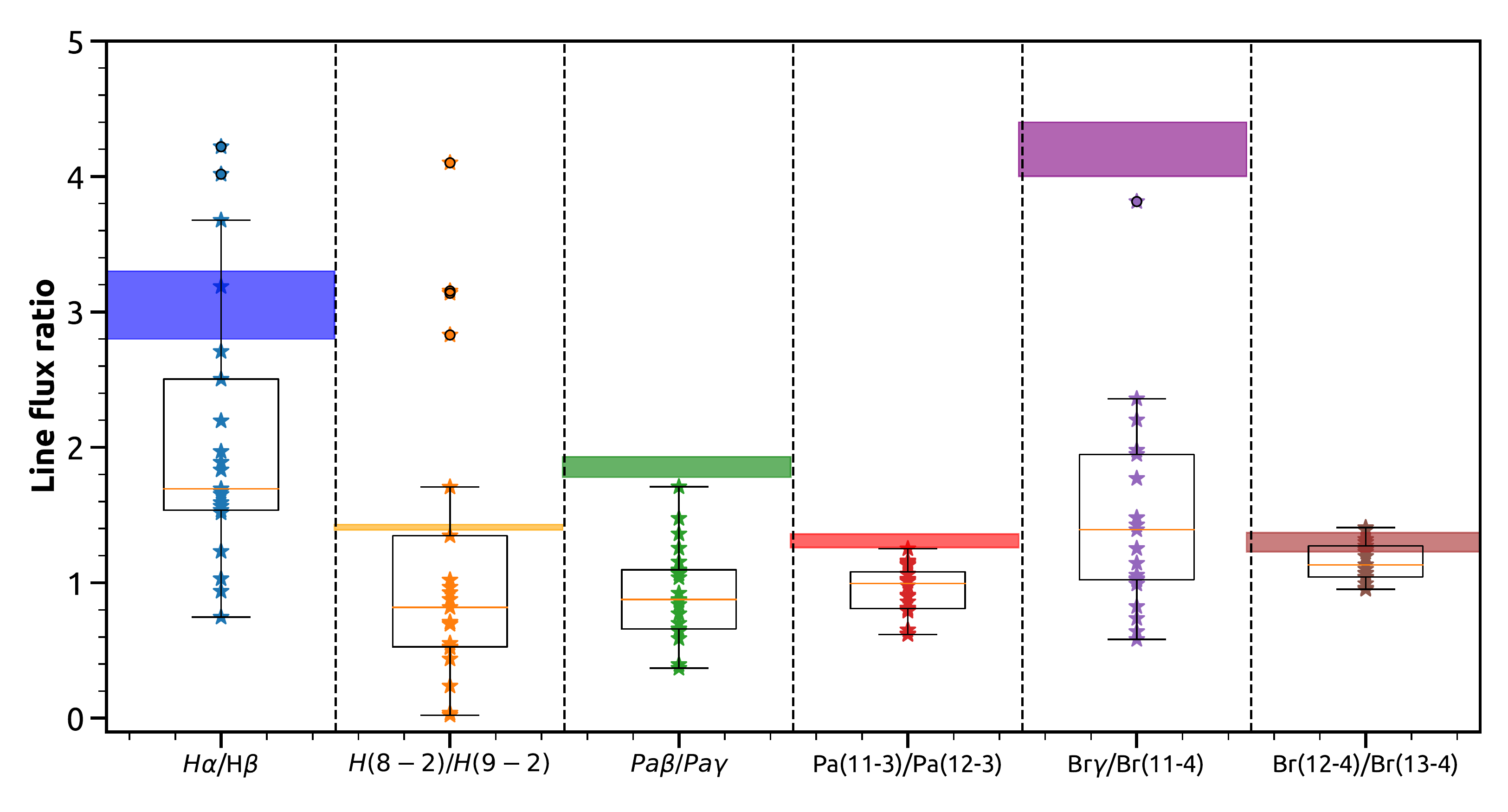}
    \caption{Boxplot representations of various H{\sc I} ratios values and the range of theoretical case B ratios are given in each case. The shaded region for each ratio denotes the range of theoretical Case B values for 2000 $<$ T $<$ 12000 and 4 $<$ log(n$_e$) $<$ 13 from \citet{storey1995recombination}.}
    \label{fig:ratio_boxplot}
\end{figure*}

\begin{figure}
    \centering
    \includegraphics[width=1\columnwidth]{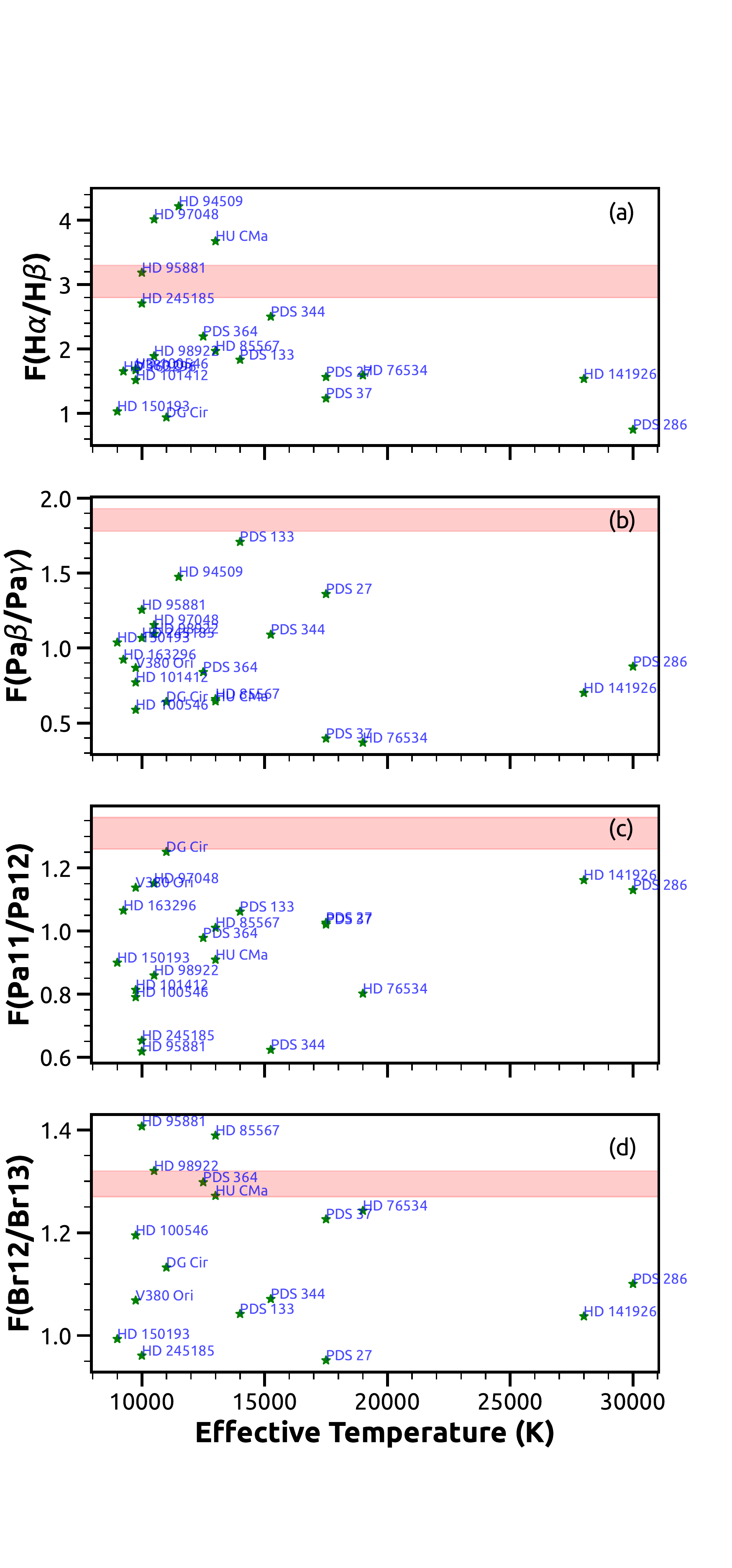}
    \caption{A scatter plot of various line flux ratios plotted against the star's effective temperature is shown. The T$_{eff}$ values are from \citet{vioque2018gaia}. The red shaded region in each panel denotes the range of theoretical Case B ratio for 2000 $<$ T $<$ 12000 and 4 $<$ log(n$_e$) $<$ 13 from \citet{storey1995recombination}. The name of the star is annotated next to each point.  }
    \label{fig:spectral_type}
\end{figure}

This work focuses on understanding the formation of H{\sc I} emission lines, especially the higher-order emission lines. A crude understanding of recombination emission tells us that the higher-order lines can appear in two conditions. One, if the ionised medium is spatially large enough or highly ionised to the extent, the higher-order transitions become favourable. Two, even if the emitting medium is small, the temperature and density of the medium play a role in modifying the emissivity ratios of higher-order lines with respect to lower-order lines.

Looking closely at the statistical results mentioned above, we see that the `Balmer I' stars show a mild correlation to the T$_{eff}$. Most HAe ($<$10,000K) belong to `Balmer III' where only H$\alpha$ is seen in emission. As we move on to higher T$_{eff}$ (late HBe stars), there is a mix of `Balmer I' and `Balmer II' stars. Thus, the T$_{eff}$ of the host star may play a role in higher-order line emission.
Further, as mentioned earlier, the older stars predominantly show only the lower-order lines in emission. However, it should also be noted that not all young stars belong to `Series I'. Several younger stars do not show higher-order lines. Under the assumption that higher-order line emission is due to circumstellar medium of the star, then the above inference shows that some younger stars lose their dynamic circumstellar environment too early in their PMS phase.

We compare the observed H{\sc I} line flux values to the theoretical Case B recombination values to understand the physical conditions from which these higher-order lines are emitted. Though it is known that Case B recombination model cannot be strictly applied to Herbig systems (owing to opacity effects and multiple H{\sc I} emission regions), we attempt to analyse the observed line ratios of `Series I' stars with the case B theoretical values.

\begin{figure*}
    \centering
    \includegraphics[width=2\columnwidth]{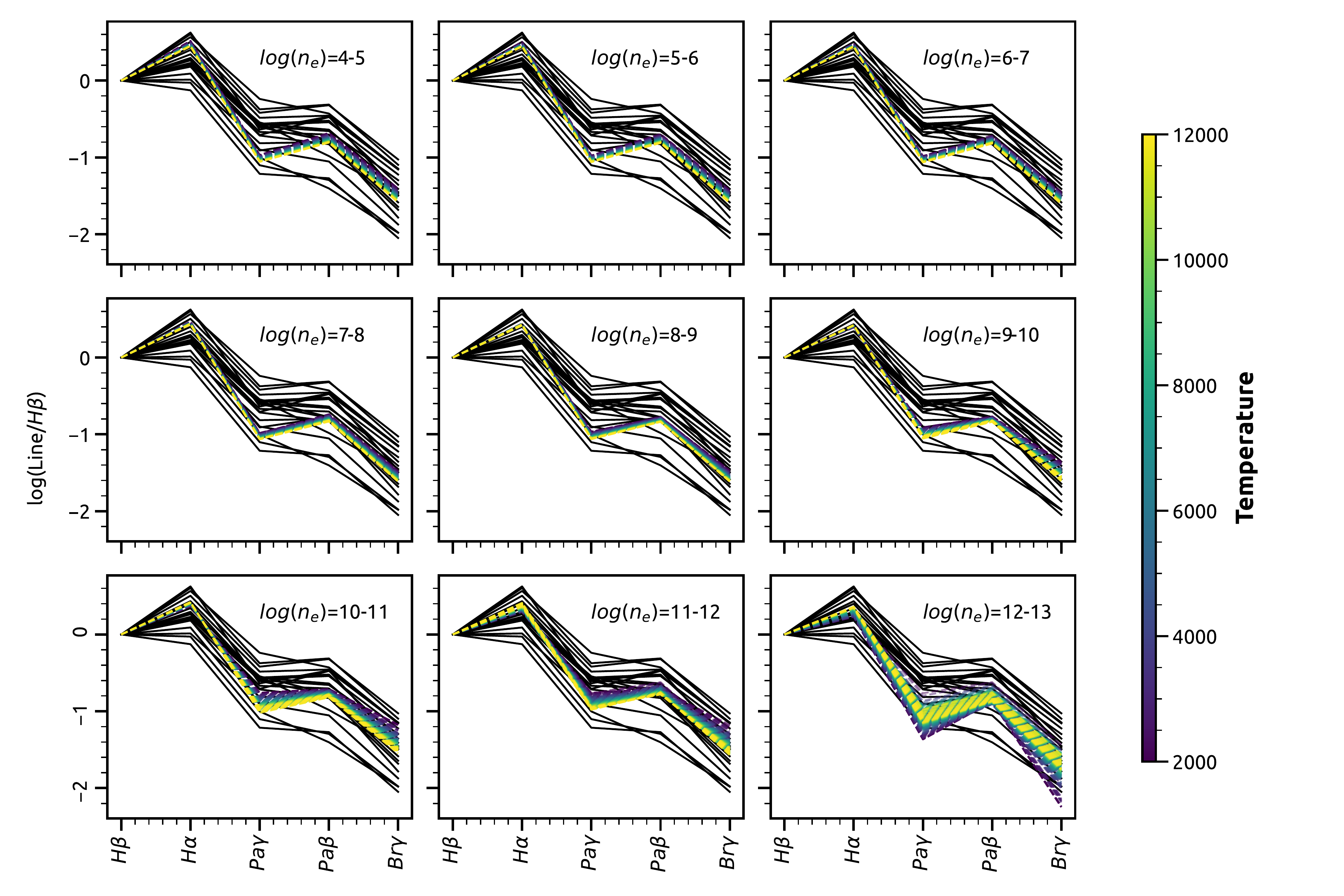}
    \caption{Case B recombination model analysis of lower order emission lines belonging to Balmer, Paschen and Brackett series. The solid black lines represent the line flux ratios of 21 HAeBe stars with lower-order lines in emission. The dotted lines in each panel represent the theoretical Case B recombination line ratios, where the colourmap denotes the corresponding temperature of emitting media.  }
    \label{fig:lower_order}
\end{figure*}

\begin{figure*}
    \centering
    \includegraphics[width=2\columnwidth]{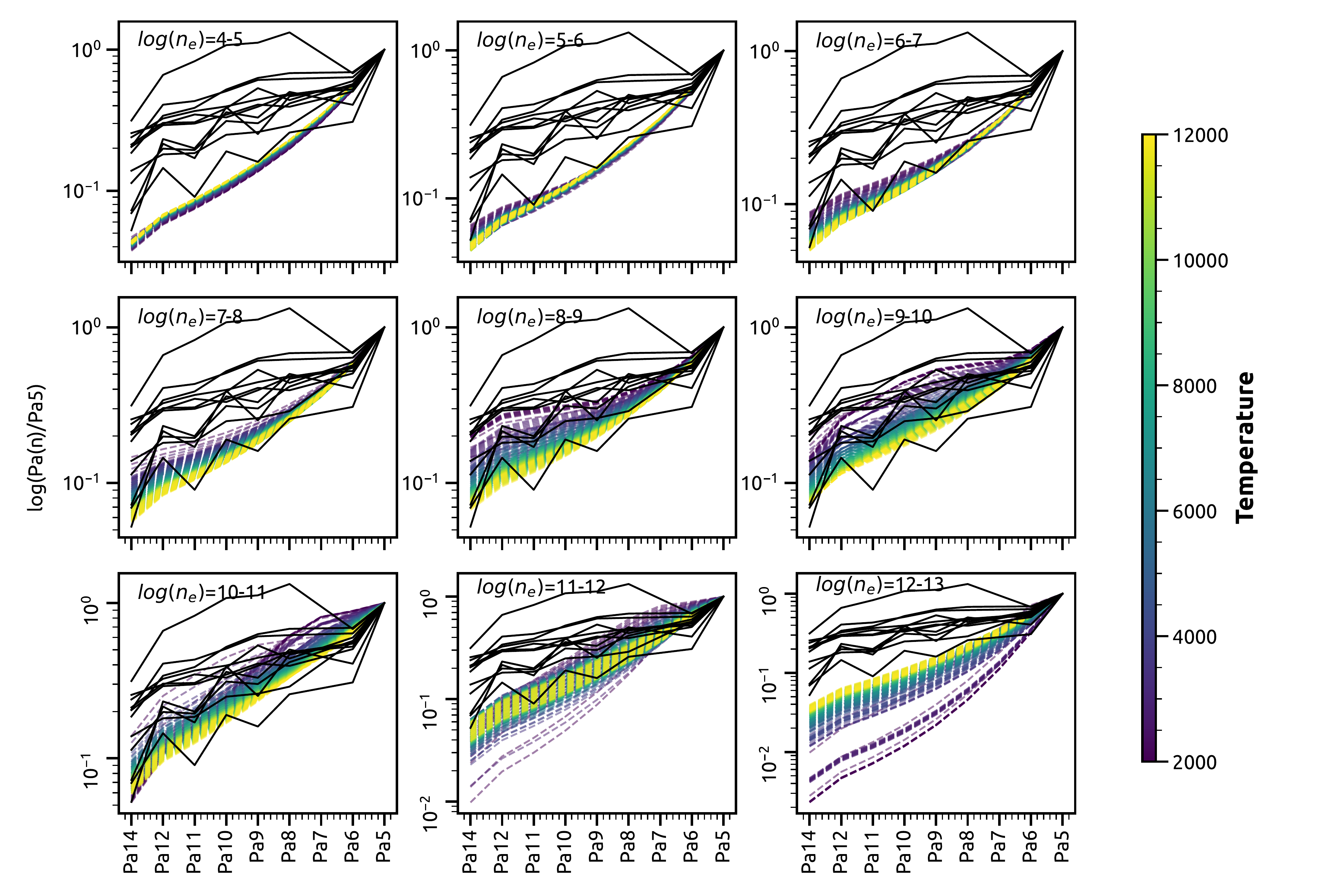}
    \caption{Case B recombination model analysis of Paschen emission lines with Pa5 as reference. The solid black lines represent the line flux ratios of 13 HAeBe stars with all higher-order lines in emission. The dotted lines in each panel represent the theoretical Case B recombination line ratios, where the colormap denotes the corresponding temperature of emitting medium. }
    \label{fig:pa_higher}
\end{figure*}

\subsection{H{\sc I} Case B recombination analysis}

\citet{baker1938physical} described two specific cases of electron recombination. Case A was applied to physical conditions where all the hydrogen transitions are optically thin, so any photons emitted by atoms within the gas escape with no subsequent interaction. Case B recombination is applied when the emitting medium is assumed to be optically thick to Lyman photons but is optically thin to photons of all other H{\sc I} transitions. \citet{hummer1987recombination} presented theoretical values of the H{\sc I} line emissivity ratio assuming different temperature and density conditions of the medium. This work compares the observed line flux ratios to different Case B recombination values based on T and log(n$_e$) as calculated by \citet{hummer1987recombination}. We use `pyneb' software to obtain Case B recombination line emissivity ratios for various temperature and density values. The Case B recombination models \citep{hummer1987recombination, storey1995recombination} provide emissivity values for H{\sc I} recombination transition for a range of temperatures (500 K $<$ T $<$ 30,000 K) and electron densities (log(n$_e$) = 2-14). 

 The EW of H{\sc I} lines used here were obtained from \citet{fairlamb2016vizier}. The EW was then converted into line fluxes using the appropriate continuum band flux. To maintain the homogeneity in obtaining continuum flux values, we used the newly released data from Gaia DR3, which using low-resolution BP/RP spectra, provides a catalogue of synthetic photometry in widely used filter passbands. The 2MASS \citep{cutri2003} J, H, and K$_S$ magnitudes were converted into flux using appropriate conversion relations to calculate NIR flux values. The broadband flux values were extinction corrected using the A$_V$ value given by \citet{vioque2018gaia}. The observed line flux for each emission line was calculated as the product of EW and the corresponding extinction corrected broadband continuum flux value \citep{mathew2018excitation}.

From 51 stars which has Gaia DR3 synthetic photometry values, we selected 21 stars which show the lower-order lines (such as H$\alpha$, H$\beta$, Pa$\beta$, Pa$\gamma$ and Br$\gamma$) in emission and of this 21 stars, 13 stars show higher-order lines of Paschen and Brackett series in emission as well. We use this sample of HAeBe stars for our following analysis. The spectral type of these stars ranges from A0 to B0, i.e, HBe stars.

As mentioned earlier, case B recombination assumes that the medium is optically thick to Lyman photons and optically thin to other photons. If this assumption is valid in the emitting medium, the $F(H\alpha/H\beta)$ ratio, for example, will stay close to the theoretical range of 2.8-3.1. However, the ratio will differ from the theoretical values if the medium becomes optically thick to other photons. Since the opacity effects are wavelength-dependent, the ratio will be less than the theoretical value for an optically thick medium. In other words, more $H\alpha$ photons will be absorbed compared to $H\beta$ photons, decreasing the $F(H\alpha/H\beta)$ ratio.

Figure \ref{fig:ratio_boxplot} shows the distribution of various H{\sc I} line ratios and the respective theoretical values. All the line ratios arise from an optically thick medium (since the ratios are less than case B values). It is interesting to note that for four stars, the Balmer line ratios ($F(H\alpha/H\beta)$ and F(H(8-2)/H(9-2))) are above the theoretical value. The cases where the ratios are higher than case B values may be explained by bringing the idea of multiple H{\sc I} emitting regions. These four stars belong to late HBe stars, namely, HD 95881 (A0), HU CMa (B7), HD 97048 (B9) and HD 94509 (B9). They can be further studied in detail to understand the additional emission component providing the line ratio excess. As we move to the Paschen series, namely, F($Pa\beta$/$Pa\gamma$) and Pa(11-3)/Pa(12-3), we see that all the stars studied here have emission medium that is optically thick to Paschen photons as well. The multiple H{\sc I} line emitting medium evident in the Balmer series is not visible in the Paschen series. This may be due to the physical conditions of the (secondary) emitting medium favouring only Balmer emission over other series. For the Brackett series, we only check the F(Br(12-4)/Br(13-4)) ratio. All the stars show values close to the optically thin case B theoretical values. This means that the medium is optically thin to Brackett series lines. It is interesting to see that the different H{\sc I} series show different characteristics. A more detailed line analysis, mainly including Pa$\alpha$, Br$\beta$ and Br$\alpha$ emission lines, is needed to conclusively show the difference of opacity effects in different series of H{\sc I}.

As we see from Figure \ref{fig:spectral_type}, there is no dependence of the line flux ratios on the effective temperature (T$_{eff}$) of the star. As the amount of ionizing flux changes rapidly with temperature, we can say that the line flux ratios do not depend on the incoming radiation. Since the spectra do not include lower-order lines of Brackett (Br$\alpha$ and Br$\beta$), comparing Brackett lower-order lines to higher-order lines is out of the scope of this work.

\begin{figure*}
    \centering
    \includegraphics[width=2\columnwidth]{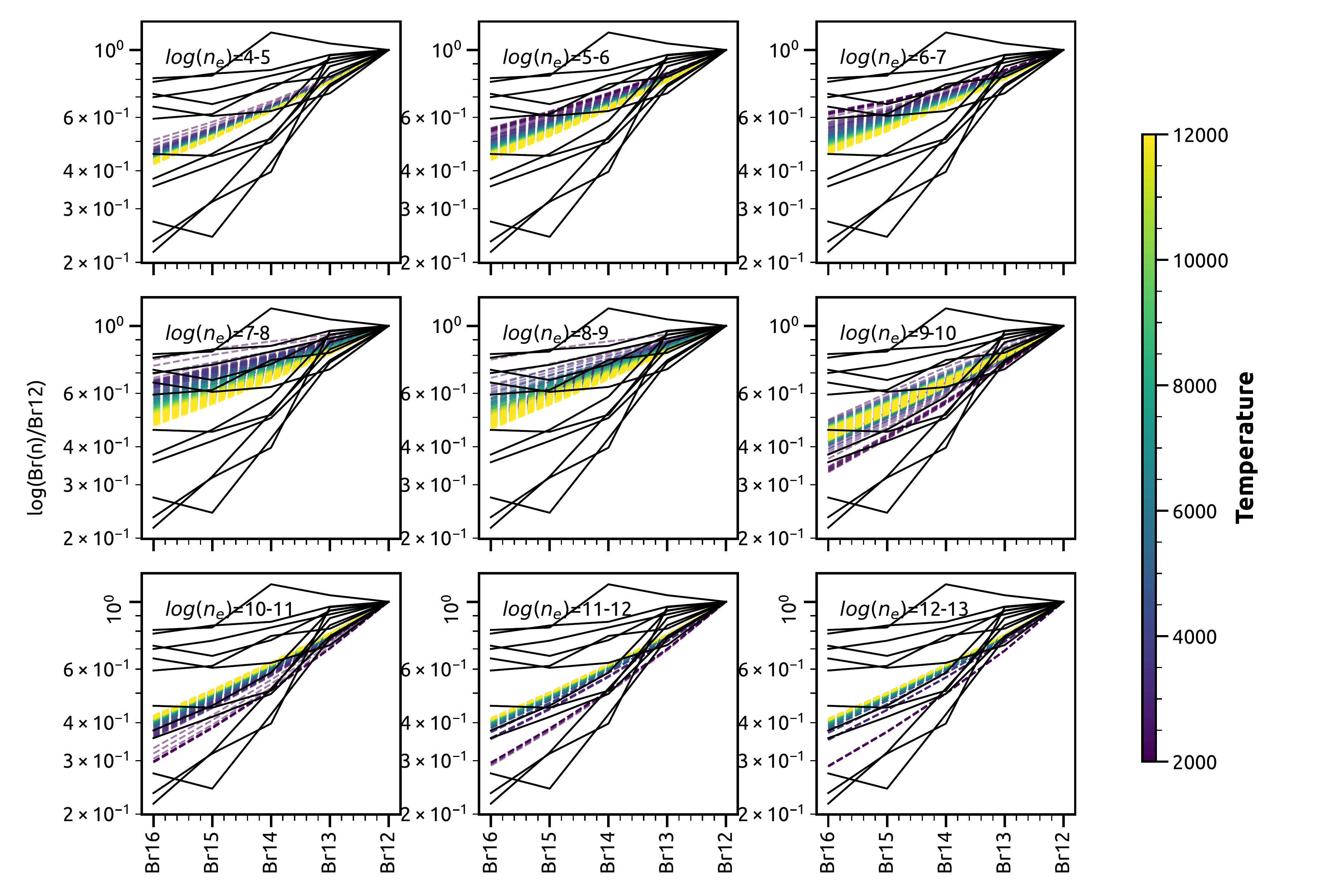}
    \caption{Case B recombination model analysis of Brackett emission lines with Br(12-5) as reference. The solid black lines represent the line flux ratios of 13 HAeBe stars with all higher-order lines in emission. The dotted lines in each panel represent the theoretical Case B recombination line ratios, where the colourmap denotes the corresponding temperature of emitting media. }
    \label{fig:br_higher}
\end{figure*}

We analysed three distinct sets of line ratios. Figures \ref{fig:lower_order} to \ref{fig:br_higher} are separated into different panels with log(n$_e$) fixed and the temperature varied. In this way, the degeneracy in the line ratios for different temperature and density values can be understood to some extent. To begin with, we used the lower-order lines present in the X-Shooter spectra for our analysis. The ratio of the lines mentioned above with respect to H$\beta$ is shown in Figure \ref{fig:lower_order}. Since H$\alpha$ is prone to opacity effects, we did not consider it our reference value. From the 21 stars used in this analysis, it can be seen that all the ratios show a range of values and do not favour any particular value. Hence, the physical properties of the emitting medium change drastically from star to star. Most of the observed Paschen decrement values match closely with theoretical values of log(n$_e$)=10-12. However, the line ratios are not sensitive to temperature to provide a range of temperature estimates. 

In addition to the lower-order lines, we also compared the ratios of Paschen higher order lines (Pa6\footnote{Pa`n' denotes the transition between upper-level n and the lower level corresponding to Paschen series. Similar convention is followed for Brackett series} to Pa14) to the theoretical Case B values. Pa13 was not considered in this analysis since it can be blended with the Ca{\sc II} triplet line at 8662$\AA$. The theoretical values of the Paschen emissivity ratio with respect to Pa5 are shown for different temperature and log(n$_e$) values. The Paschen decrement does not seem sensitive to temperature for log(n$_e$) values between 4-7. As the electron density decreases, the Paschen decrement values become more and more temperature sensitive, especially in the higher-order lines of Paschen (Pa8 to Pa14). In this analysis, we find that most of the stars match with log(n$_e$) = 8-11 values with temperature values of around 3000-6000K (2nd row of Figure \ref{fig:pa_higher}). The Case B density values observed here are less than the best-matched values obtained using lower-order line ratios.

The above analysis was repeated with Brackett higher order lines (Br12-Br16). We did not consider Br17-Br20 because they are very weak, and the values may be erratic. The theoretical values of the Brackett decrement seem to be more temperature sensitive than the Paschen decrement for log(n$_e$) values between 7 and 10. Similar to what was observed in Figure \ref{fig:pa_higher}, most stars match with log(n$_e$) = 7-11. It can also be seen that the ratio of the Brackett series is more sensitive to density and temperature values compared to the Paschen series. 

The above analysis shows that the H{\sc I} line ratios do not follow a particular trend, especially since they do not match the theoretical case B values. A different model for recombination, like the \citet{kwan2011origins}, applicable for winds and accretion flows in Classical T Tauri stars, should be tested for line ratio analysis in HAeBe systems as well. Interestingly, the observed Brackett series ratio varies drastically from star to star compared to the Paschen series ratio. In this work, we are limited by the availability of high signal-to-noise ratio spectra, which will help constrain higher-order lines' EW ratios more accurately.

\section{Conclusion}

We carried out a spectroscopic analysis of a subset of HAeBe stars using the archival X-Shooter spectra. X-Shooter provides a unique opportunity by providing a near-simultaneous wavelength coverage starting from 3000 $\AA$ to 25000 $\AA$, which aids in measuring the emission lines of the H{\sc I} series in the same epoch. As an initial step into understanding the line ratios of different H{\sc I} emission lines, we queried the pre-existing X-Shooter spectra of well-characterised HAeBe stars and visually checked for the presence of various H{\sc I} lines. We visually classified the Herbig stars into different categories based on the presence of H{\sc I} emission lines. These segregated groups of stars are checked for a correlation between the presence of higher-order H{\sc I} lines and stellar parameters. 

From our analysis of stellar parameters belonging to different groups, it can be seen that most of the stars massive than 10 M$_\odot$ show all the series lines in emission. The presence of higher-order emission lines in most HBe stars points to the distinction in the circumstellar environment and change in accretion mode between HAe and HBe stars. As expected, most stars showing all emission lines are very young ($<$2 Myr). There is no clear dependence of the presence of series lines on the star's effective temperature. However, as mass and spectral type of star are correlated, `Series I' stars primarily belong to B spectral type, and `Series II/III' stars mainly belong to the A spectral type. `Series I' stars show intense ($>$ 50 $\AA$) H$\alpha$ emissions when compared to `Series II' (20-60 $\AA$) and `Series III' (5-20 $\AA$) HAeBe stars. 

Then we performed group-as-whole Case B recombination analysis of 21 stars in 'Series I and II'. We see that the line flux ratios do not show any dependence on the star's spectral type, which points to the idea that the ratios do not depend on the incoming ionising radiation. Based on the trend of line ratios shown in Figure \ref{fig:ratio_boxplot} and Figure \ref{fig:spectral_type}, we see that the optical depth of the medium affects various H{\sc I} series differently. An in-depth analysis of opacity dependence of wavelength is required to understand this further. The line flux ratios of lower order lines belonging to Balmer, Paschen and Brackett are compared to various temperature and log(n$_e$) Case B ratios. We find that, even though the temperature cannot be adequately constrained using these ratios, the line ratios of most stars closely matches for $log(n_{e})$ = 9-11. Then we analysed the Paschen line decrements with Pa(5-3) as the reference. This analysis, in addition to validating the log(n$_e$) = 9-11, also provides an estimate of temperature in the range 3000 - 6000 K compared to values provided by \citet{bary2008quiescent} for TTS. The same analysis is repeated using the higher order Brackett lines where most line ratios do not match any of the Case B recombination models. It might be due to the weak Br14-16 lines where the EW errors are sometimes similar to observed EW values. We plan to study the Brackett series in more detail using the infrared spectroscopic surveys such as APOGEE \citep{apogee2007}. 

\section*{Acknowledgements}

We want to thank the Science \& Engineering Research Board (SERB), a statutory body of the Department of Science \& Technology (DST), Government of India, for funding our research under grant number CRG/2019/005380. RA acknowledges the financial support from SERB POWER fellowship grants SPF/2020/000009. The authors are grateful to the Centre for Research, CHRIST (Deemed to be University), Bangalore, for the research grant extended to carry out the current project (MRPDSC-1932). We thank the SIMBAD database and the online VizieR library service for helping us with the literature survey and obtaining relevant data. This work has made use of data from the European Space Agency (ESA) mission {\it Gaia} (\url{https://www.cosmos.esa.int/gaia}), processed by the {\it Gaia} Data Processing and Analysis Consortium (DPAC, \url{https://www.cosmos.esa.int/web/gaia/dpac/consortium}). Funding for the DPAC has been provided by national institutions, in particular, the institutions
participating in the {\it Gaia} Multilateral Agreement. 

\balance
\bibliography{biblio}

\end{document}